\documentstyle[12pt,fleqn]{article}

\evensidemargin=\oddsidemargin
\begin{document}
\bigskip
\begin{center}
\begin{Large}
\begin{bf}
The Magnetic Behavior of the t-J Model
\end{bf}
\end{Large}
\bigskip
\bigskip
$\;$\\
{\centerline {Yu-Liang Liu}}
International Centre for Theoretical Physics, P. O. Box 586, 34100
Trieste, Italy\\
\bigskip
{\bf Abstract}
\end{center}
\begin{center}
\begin{minipage}[c]{13cm}
\begin{sf}

\parindent=1.2cm
Using the spin-hole coherent state representation, and taking the long
range antiferromagnetic N\'{e}el order as the background of the spin
degree part, we have studied the magnetic behavior of the t-J model in
the usual slave boson and slave fermion treatment of the single
occupation constraint, and shown that we can qualitatively explain the
anomalous magnetic and transport properties of the normal state of the
cuprate superconducting materials by the t-J model.

\end{sf}
\end{minipage}
\end{center}
\newpage

Recently, the significant progress has been made in the understanding of
the low energy spin dynamics of the normal state of the cuprate
superconducting materials in both theoretical$^{[1-4]}$ and
experimental$^{[5-7]}$ aspects. In the undoping case, the spin dynamics
of the cuprates, such as $La_{2}CuO_{4}$, is well described by the
quantum Heisenberg model on a square lattice of $Cu$ sites. The authors
of Refs.[1,2] have extensively studied it by using the scalling and
renormalization group theory and /or large-N expansion methods, and have
given some valuable results which are in good agreement with the current
experimental data. However, in the doped case, up to now there is not a
general consensus on choosing a microscopic theory qualitatively to
describe the unusually magnetic and transport properties of the normal
state over the entire doping range from insulator to high doped
compounds, although many models have been proposed to describe them.

The unusually physical properties of the normal state of the cuprate
superconducting materials may originate from their strongly
antiferromagnetic correlation. The doping will destroy the long range
antiferromagnetic correlation, but the system still maintains a strongly
short range antiferromagnetic correlation. In Ref. 4, we have given a
detail study following this idea, and obtained some results which can
qualitatively explain the unusually physical properties of the normal
state. In this paper, using the similar method as in Ref. 4, we study the
magnetic behavior of the t-J model. It is well-known that the gauge
theory of the t-J model$^{[8-10]}$ gives a better description of the
transport property of the normal state, but up to now one has not known
whether it can also give a reasonable description to the magnetic
behavior of the normal state.

We first adopt an usual method to deal with the single occupation
condition by introducing a slave fermion, so the Hamiltonian of the t-J
model can be written as in a hole representation
\begin{equation}\begin{array}{rl}
H= & t\displaystyle{\sum_{<ij>}}(f^{+}_{j}f_{i}b^{+}_{i\sigma}b_{j\sigma}
+h.c)\\
& +J\displaystyle{\sum_{<ij>}}(1-f^{+}_{i}f_{i})\hat{S}_{i}\cdot\hat{S}_{j}
(1-f^{+}_{j}f_{j})+\displaystyle{\sum_{i}}\lambda_{i}(1-f^{+}_{i}f_{i}
-b^{+}_{i\sigma}b_{i\sigma})
\end{array}\end{equation}
where $\hat{S}_{i}=\frac{1}{2}b^{+}_{i\alpha}\hat{\sigma}_{\alpha\beta}
b_{i\beta}$, $b_{i\sigma}$ is a hard-core boson operator which describes
the spin degree of the electron, and $f_{i}$ is a fermion operator which
describes the charge degree of the electron. The electron operator is
$c_{i\sigma}=f^{+}_{i}b_{i\sigma}$, $\lambda_{i}$ is a Lagrangian
multiplier which ensures the single occupation condition of the
electrons. In the spin-hole coherent state representation introduced by
Auerbach$^{[11]}$
\begin{equation}
|\hat{\Omega}, \xi>_{S}\equiv |\hat{\Omega}>_{S}\otimes |0>_{f}+
|\hat{\Omega}>_{S-\frac{1}{2}}\otimes\xi f^{+}|0>_{f}
\end{equation}
where $|\hat{\Omega}>_{S}$ is a spin coherent state$^{[12]}$ and $\xi$ is
an anticommuting Grassmann variable, the partition functional of the
Hamiltonian (1) can be written as
\begin{equation}
Z=\int D\hat{\Omega}D\xi^{*}D\xi exp\{-\int^{\beta}_{0}[{\it L}_{\Omega}+
{\it L}_{\xi}]\}
\end{equation}
\begin{equation}
{\it L}_{\Omega}=-i\sum_{i}2S\omega_{i}+JS^{2}\sum_{<ij>}(1-\xi^{*}_{i}
\xi_{i})\hat{\Omega}_{i}\cdot\hat{\Omega}_{j}(1-\xi^{*}_{j}\xi_{j})
\end{equation}
\begin{equation}
{\it L}_{\xi}=\sum_{i}\xi^{*}_{i}(\partial_{\tau}+i\omega_{i}+\mu_{i})
\xi_{i}+\sqrt{2}tS\sum_{<ij>}(\xi^{*}_{j}\xi_{i}e^{i\gamma_{ij}}\sqrt{
1+\hat{\Omega}_{i}\cdot\hat{\Omega}_{j}}+h.c)
\end{equation}
where the Berry phase $\omega$ is a functional of the spin order
parameter $\hat{\Omega}(\tau)$. It is ambiguous modulo $4\pi$, and its
functional derivative is quite well-behaved$^{[12]}$
\begin{equation}
\int d\tau\delta\omega=\int d \tau\hat{\Omega}\cdot(\partial_{\tau}
\hat{\Omega}\times\delta\hat{\Omega})
\end{equation}
The parameter $\mu_{i}$ is a chemical potential of the slave fermion
$\xi$, $\gamma_{ij}$ is the phase factor of
$_{S}<\hat{\Omega}|b^{+}_{i\sigma}b_{j\sigma}|\hat{\Omega}>_{S}$.
The Lagrangian ${\it L}_{\xi}$ is invariant under following gauge
transformations
\begin{equation}
\xi_{i}\rightarrow\xi_{i}e^{i\theta_{i}},\;\;
\gamma_{ij}\rightarrow\gamma_{ij}-\theta_{i}+\theta_{j},\;\;
\mu_{i}\rightarrow\mu_{i}+i\partial_{\tau}\theta_{i}
\end{equation}
which derives from the slave fermion representation of the electron
operator $c_{i\sigma}=f^{+}_{i}b_{i\sigma}$. The single occupation
condition in (1) disappears in (4) and (5), because in the spin-hole
coherent state representation the term
$(1-f^{+}_{i}f_{i}-b^{+}_{i\sigma}b_{i\sigma})$ is equal to zero at each
site. From the equations (4) and (5), we see that the Lagrangian ${\it
L}_{\Omega}$ dominates the antiferromagnetic behavior of the system, then
the Lagrangian ${\it L}_{\xi}$ dominates the ferromagnetic behavior (or
destroys the antiferromagnetic behavior) of the system because the
factor $\sqrt{1+\hat{\Omega}_{i}\cdot\hat{\Omega}_{j}}$ is zero for
antiferromagnetic order and is biggest for ferromagnetic order. According
to the current experimental data of the cuprate superconducting
materials, almost all of them show a strongly short range
antiferromagnetic behavior in the normal state, even in the
superconducting state, the short range antiferromagnetic behavior also
appears. Therefore, according to this fact, we take a long range
antiferromagnetic N\'{e}el order as a background of the spin order parameter
\begin{equation}
\hbar S\hat{\Omega}_{i}\simeq\hbar\eta_{i}\hat{\Omega}(x_{i})
+a^{2}\hat{L}(x_{i})
\end{equation}
where $a^{2}$ is the unit cell volume, $\hat{\Omega}(x_{i})$ is the
slowly varying N\'{e}el unit vector order, i.e., spin parameter field
$|\hat{\Omega}(x_{i})|=1$, and $\hat{L}(x_{i})$ is the slowly varying
magnetization density field, $\hat{\Omega}(x_{i})\cdot\hat{L}(x_{i})=0$.
The Berry phase term may be separated into two parts
\begin{equation}
S\sum_{i}\omega_{i}\simeq S\sum_{i}\eta_{i}\omega(x_{i})+
\frac{1}{\hbar}\int
d^{2}x\hat{\Omega}\cdot(\frac{\partial\hat{\Omega}}
{\partial\tau}\times\hat{L})
\end{equation}
where $\omega(x)$ is the solid angle subtended on the unit sphere by the
closed curve $\hat{\Omega}(x,\tau)$ (parametrized by $\tau$). Because of
in the long range antiferromagnetic N\'{e}el order approximation, the
electron hoping must be accompanied with a $\pi-phase$ rotation in spin
space to match with the nextest neighbor spin orientations, so the t-term
in (1) must be changed as
\begin{equation}\begin{array}{rl}
f^{+}_{i}f_{j}b^{+}_{j\sigma}b_{i\sigma}=&
\displaystyle{e^{-2i\displaystyle{\sum_{l\neq
i,j}\theta_{ij}(l)S^{z}_{l}}}}f^{+}_{i} f_{j}\displaystyle{
e^{2i\displaystyle{\sum_{l\neq i,j}\theta_{ij}(l)S^{z}_{l}}}}
b^{+}_{j\sigma}b_{i\sigma}\\
=&  \displaystyle{e^{-2i\displaystyle{\sum_{l\neq i,j}\theta_{ij}(l)
S^{z}_{l}}}}f^{+}_{i}f_{j}\tilde{b}^{+}_{j\sigma}\tilde{b}_{i\sigma}
\end{array}\end{equation}
where $\theta_{ij}(l)=\theta_{i}(l)-\theta_{j}(l)$, $\theta_{i}(l)$ is an
angle between the direction from site $i$ to site $l$ and some fixed
direction, the $x$ axis for example; $S^{z}_{l}=\frac{1}{2}b^{+}_{l\alpha}
\sigma^{z}_{\alpha\beta}b_{l\beta}$, the $z$-component of the spin
operator; $\tilde{b}_{i\sigma}=e^{2i\sum_{l\neq
i}\theta_{i}(l)S^{z}_{l}}b_{i\sigma}$, is a fermion operator.
Under the approximations (8) and (9), and eliminated the magnetization
density field $\hat{L}(x)$, the Lagrangians in (4) and (5) can be written
as, respectively
\begin{equation}
{\it L}_{\Omega}=\frac{1}{2g_{0}}\int
d^{2}x[(\vec{\partial}\hat{\Omega})^{2}+\frac{1}{c^{2}}(\partial_{\tau}
\hat{\Omega})^{2}]
\end{equation}
\begin{equation}\begin{array}{rl}
{\it L}_{\xi}=& \displaystyle{\sum_{i}}\xi^{*}_{i}(\partial_{\tau}-\mu_{i})
\xi_{i}\\
+&  \sqrt{2}tS\displaystyle{\sum_{<ij>}\{\xi^{*}_{j}\xi_{i}
e^{i\gamma_{ij}^{'}}[1+\eta_{i}\eta_{j}\hat{\Omega}(x_{i})\hat{\Omega}
(x_{j})]^{\frac{1}{2}}}+h.c\}
\end{array}\end{equation}
where $\gamma^{'}_{ij}=\gamma_{ij}+\displaystyle{\sum_{l\neq i,j}}
\theta_{ij}(l)_{S}<\hat{\Omega}|(b^{+}_{l\uparrow}b_{l\uparrow}-
b^{+}_{l\downarrow}b_{l\downarrow})|\hat{\Omega}>_{S}$,
$g_{0}=(J(1-\delta)^{2}S^{2})^{-1}$, $c^{2}=8(aJ(1-\delta)S)^{2}$. For the
$J$-term in (4), we have replaced the $f^{+}_{i}f_{i}$ and
$f^{+}_{j}f_{j}$ by $\delta=<f^{+}_{i}f_{i}>=<f^{+}_{j}f_{j}>$, the
doping density. We have omitted the terms $\sum_{i}\eta_{i}\omega(x_{i})$
and $\sum_{i}\eta_{i}\omega(x_{i})\xi^{*}_{i}\xi_{i}$. If $\omega(x)$ is
a slowly varying function of space coordinates $\vec{x}$ and "time"
$\tau$ and the occupation number of the quasiparticle $\xi$ is equal at
the even and odd sites, these two terms have a little contribution to the
system. However, the quantity $\omega(x)$ provides an attractive
interaction between the fermions $\xi_{i}$ and $\xi_{i+\hat{\delta}}$,
$\hat{\delta}=(\pm a, \pm a)$, at the even and odd sites, respectively,
which may induce the pairing between the slave fermions at the even and odd
sites. Here we assume this effect is very small, and do not consider it,
or we only consider the normal state of the system.

Taking the Hartree approximation, the Lagrangian (12) can be written as
\begin{equation}\begin{array}{rl}
{\it L}_{\xi}=&  \displaystyle{\sum_{i}}\xi^{*}_{i}(\partial_{\tau}-
\mu_{i})\xi_{i}+tS\displaystyle{\sum_{<ij>}}\{\Delta_{ij}\xi^{*}_{j}
\xi_{i}\\
+&
\Lambda_{ij}\xi^{*}_{j}\xi_{i}
[(\hat{\Omega}(x_{i})-\hat{\Omega}(x_{j}))^{2}]^{1/2}
+\Gamma_{ij}e^{i\gamma^{'}_{ij}}+h.c\}
\end{array}\end{equation}
where $\Delta_{ij}=<[(\hat{\Omega}(x_{i})-\hat{\Omega}(x_{j})^{2}]^{1/2}e^{
i\gamma^{'}_{ij}}>$, $\Lambda_{ij}=<e^{i\gamma^{'}_{ij}}>$, $\Gamma_{ij}=
<\xi^{*}_{j}\xi_{i}[(\hat{\Omega}(x_{i})-\hat{\Omega}(x_{j}))^{2}]^{1/2}$.
The $\Delta_{ij}$ term is the kinetic term of the slave fermion $\xi$,
the $\Lambda_{ij}$ term is the interaction term between the slave fermion
$\xi$ and the spin parameter field $\hat{\Omega}(x)$, the $\Gamma_{ij}$
term is a gauge field Maxwell-like term which provides a background gauge
current to maintain the system being neutral to the native gauge field
deriving from the slave fermion representation of the electron (see
below). To retain the gauge invariance of the Lagrangian (13) under the
gauge transformations in (7), we can take following approximations
\begin{equation}
\Delta_{ij}=\Delta e^{i\theta_{ij}},\;\;
\Lambda_{ij}=\Lambda e^{i\theta_{ij}},\;\;
\Gamma_{ij}=\Gamma e^{-i\theta_{ij}}
\end{equation}
where
$\theta_{ij}=(\vec{x}_{i}-\vec{x}_{j})
\cdot\vec{A}(\frac{x_{i}-x_{j}}{2})$.
In the continuous limit, the Lagrangian (13) can be written as
\begin{equation}\begin{array}{rl}
{\it L}_{\xi}=& \int d^{2}x[\xi^{*}(\partial_{\tau}-iA_{0})\xi+
\frac{1}{2m}\xi^{*}(\vec{\partial}-i\vec{A})^{2}\xi\\
+& V\xi^{*}\xi|\vec{\partial}\hat{\Omega}|+\alpha F[a+A]]
\end{array}\end{equation}
where $m=(2tS\Delta)^{-1}$, $V=2tS\Lambda/a$, $\alpha=2tS\Gamma$,
$\mu_{i}=iA_{0}$, $|\vec{\partial}\hat{\Omega}|\equiv|(\partial_{x}
\hat{\Omega})^{2}|^{1/2}+|(\partial_{y}\hat{\Omega})^{2}|^{1/2}$,
$(\vec{x}_{i}-\vec{x}_{j})\cdot\vec{a}(\frac{x_{i}-x_{j}}{2})\simeq
-\gamma_{ij}-\displaystyle{\sum_{l\neq i,j}}\theta_{ij}(l)_{S}<
\hat{\Omega}|b^{+}_{l\uparrow}b_{l\uparrow}-b^{+}_{l\downarrow}
b_{l\downarrow}|\hat{\Omega}>_{S}$, $\vec{a}$ is a gauge field deriving
from the spin fluctuation. In the first order approximation, we can take
$F[A]$ as $F[A]=F^{2}_{ij}[a]$, $F_{ij}[A]=\partial_{i}A_{j}-
\partial_{j}A_{i}$. From the equation (15), we have the neutral condition
of the system
\begin{equation}
-\vec{j}_{\xi}+\alpha\frac{\delta F[a+A]}{\delta\vec{A}}=0
\end{equation}
The term $\alpha F[a+A]$ in (15) is similar to the gauge fluctuation term
obtained by integrating out the spin degree $b_{i\sigma}$ in the ordinary
slave fermion method of the t-J model.

We can also adopt the slave boson method to deal with the t-J
model. In the long range antiferromagnetic N\'{e}el order approximation,
to match with the nextest neighbor spin orientations, the $t$-term can be
written as
\[\begin{array}{rl}
b^{+}_{j}b_{i}f^{+}_{i\sigma}f_{j\sigma}=&
e^{-2i\displaystyle{\sum_{l\neq i,j}\theta_{ij}(l)S^{z}_{l}}}b^{+}_{j}
b_{i}e^{2i\displaystyle{\sum_{l\neq i,j}\theta_{ij}(l)S^{z}_{l}}}
f^{+}_{i\sigma}f_{j\sigma}\\
=&  e^{-2i\displaystyle{\sum_{l\neq i,j}\theta_{ij}(l)S^{z}_{l}}}
e^{-i\displaystyle{\sum_{l\neq i,j}\theta^{'}_{ij}(l)b^{+}_{l}b_{l}}}
\tilde{b}^{+}_{j}\tilde{b}_{i}\tilde{f}^{+}_{i\sigma}
\tilde{f}_{j\sigma}
\end{array}\]
where, $S^{z}_{l}=\frac{1}{2}f^{+}_{l\alpha}\sigma^{z}_{\alpha\beta}
f_{l\beta}$, $\tilde{f}_{i\sigma}=e^{-2i\displaystyle{\sum_{l\neq
i}\theta_{i}(l)S^{z}_{l}}}f_{i\sigma}$ is a hard-core boson;
$\tilde{b}_{i}=e^{i\displaystyle{\sum_{l\neq i}\theta^{'}_{i}(l)
b^{+}_{l}b_{l}}}b_{i}$ is a fermion; because of in the slave boson
representation,
the electron operator reads $c_{i\sigma}=b^{+}_{i}f_{i\sigma}$, $b_{i}$
is a slave (hard-core) boson, $f_{i\sigma}$ is a fermion. In order to use
the spin-hole coherent state representation, we have changed the slave
boson into a fermion. After finished the calculation, we can change the
fermion $\tilde{b}_{i}$ into the slave boson $b_{i}$, so we can obtain
following Lagrangian similar to that in (15)
\begin{equation}\begin{array}{rl}
{\it L}_{\bar{\xi}}=&
\int d^{2}x[\bar{\xi}^{*}(\partial_{\tau}-iA_{0})\bar{\xi}+
\frac{1}{2m^{'}}\bar{\xi}^{*}(\vec{\partial}-i\vec{A})^{2}\bar{\xi}\\
+& V^{'}\bar{\xi}^{*}\bar{\xi}|\vec{\partial}\hat{\Omega}|+
\alpha^{'} F[a^{'}+A]]
\end{array}\end{equation}
where $(\vec{x}_{i}-\vec{x}_{j})\cdot\vec{a}^{'}=-\bar{\gamma}_{ij}-
\displaystyle{\sum_{l\neq i,j}}\theta_{ij}(l)_{S}<\hat{\Omega}|
\tilde{f}^{+}_{l\uparrow}\tilde{f}_{l\uparrow}-\tilde{f}^{+}_{l\downarrow}
\tilde{f}_{l\downarrow}|\hat{\Omega}>_{S}$, $\bar{\gamma}_{ij}$ is the
phase factor of $_{S}<\hat{\Omega}|\tilde{f}^{+}_{i\sigma}
\tilde{f}_{j\sigma}|\hat{\Omega}>_{S}$, $\bar{\xi}$ is a hard-core boson
field.

If we integrate out the slave fermion $\xi$, we can obtain following
effective actions of the gauge field and the spin parameter field
\begin{equation}
S_{eff.}=S_{eff.}^{\xi}[A]+S^{\xi}_{eff.}[\Omega]
\end{equation}
\begin{equation}\begin{array}{rl}
S^{\xi}_{eff.}[A]=& \displaystyle{\int\frac{d\omega}
{2\pi}\int\frac{d^{2}q}{(2\pi)^{2}}
(\chi_{F}q^{2}-\frac{i\omega}
{v_{F}q})(\delta_{ij}-\frac{q_{i}q_{j}}{
q^{2}})A_{i}(q,\omega)A_{j}(-q,-\omega)}\\
+& \displaystyle{\int d\tau\int d^{2}x \alpha F[a+A]}
\end{array}\end{equation}
\begin{equation}
S^{\xi}_{eff.}[\Omega]=-\beta\sum_{n}\int\frac{d^{2}q}{(2\pi)^{2}}
\frac{|\omega_{n}|}{\omega_{F}}|\hat{\Omega}|^{2}(q,\omega_{n})
\end{equation}
where $\chi_{F}$ is the diamagnetic susceptibility of the slave fermion
system, and $v_{F}$ is the Fermi velocity of the slave fermion,
$\omega_{F}\propto\frac{1}{V^{2}k_{F}}$, a character energy scale which
describes the damping of the quasiparticle-hole pairing excitation to the
spin wave spectrum. We have omitted the crossover terms between the gauge
field $\vec{A}$ and the spin parameter field $\hat{\Omega}$ that are
higher order terms, and the higher order derivatives of the spin
parameter field $\hat{\Omega}$. We must be careful in dealing with the
Lagrangian (17), because as we integrate out the hard-core boson field
$\bar{\xi}$, we meet with a problem of the condensation of the hard-core
boson, only as the temperature $T>T_{0}$, the condensation temperature of
the hard-core boson, we can obtain an effective action similar to that in
equations (19) and (20). The condensation of the hard-core boson will
destroy the gauge invariance of the native gauge field $\vec{A}$.
Mathematically, the Lagrangians in (15) and (17) are equivalent, so we
think that the quantum gauge fluctuation can drastically suppress the
condensation of the hard-core bosons, if there is not another interaction
source, the condensation temperature tends to zero, $T_{0}=0$.

{}From equations (11) and (20), we obtain an effective action of the spin
parameter field of the t-J model
\begin{equation}
S_{eff.}[\Omega]=\beta\sum_{n}\int\frac{d^{2}q}{(2\pi)^{2}}\{
\frac{1}{2g_{0}}(q^{2}+\frac{1}{c^{2}}\omega^{2}_{n})-\frac{|
\omega_{n}|}{\omega_{F}}\}|\Omega|^{2}(q,\omega_{n})
\end{equation}
where $|\Omega(x,\tau)|=1$, the origin points of $q$ are in the corner
points $\vec{Q}=(\pm\frac{\pi}{a},\pm\frac{\pi}{a})$. The action (21),
our central result in this paper, is the same as that in Ref.4 we
obtained from a p-d model or an effective Hamiltonian derived from a
three-band Hubbard model. This action has two critical regions: one is a
$z=1$ (where $z$ is a dynamic exponent) region which is consisted of
three regimes: a renormalized classical (RC) regime, a quantum critical
(QC) regime and a quantum disorder (QD) regime$^{[1]}$; another one is a
$z=2$ region which maybe is also divided into the same two (QC and QD)
regimes as above, but their behavior is completely different from that in
the $z=1$ region. In the undoping case, $\omega_{F}\rightarrow\infty$,
the system is in the RC regime$^{[1,2]}$. In the underdoping case,
$\omega_{c}<\omega_{F}<\infty$, the system is in the $z=1$ QC and/or QD
regimes$^{[3,4]}$. In the optimal doping case, $\omega_{F}<\omega_{c}$,
the system goes into the $z=2$ region$^{[3,4]}$. $\omega_{c}$ is a
characteristic energy scale which indicases a crossover of the system
from the $z=1$ region to the $z=2$ region as doping.
We see that the $\omega_{F}$ term in (21) which derives from the
damping of the quasiparticle-hole pairing excitation to the spin wave
spectrum is very important for determining the doping influence on the
system, especially in the optimal doping case, this term is dominant.

Generally, in the $z=1$ region, the $\omega_{F}$ term is very small, and
can be treated perturbatively, in the low energy limit we can obtain
following spin susceptibility
\begin{equation}
\chi(q,\omega)=\frac{\chi_{0}}{\xi^{-2}+q^{2}-\frac{1}{c^{2}}\omega^{2}
-\frac{i\omega}{\omega^{R}_{F}}}
\end{equation}
where $\xi$ is a coherent length, $\omega^{R}_{F}$ is a renormalized
characteristic energy scale of the spin fluctuation. In the ($z=1$) QC
regime$^{[4]}$, $\xi\sim\frac{1}{T}, \omega^{R}_{F}\sim\frac{1}{T}$; In
the ($z=1$) QD regime, $\xi$ and $\omega^{R}_{F}$ take constants. In the
$z=2$ region, the $\omega_{F}$ term is dominant, the $\omega^{2}$ term is
irrelevant and can be omitted, in the low energy limit we can obtain
following spin susceptibility
\begin{equation}
\bar{\chi}(q,\omega)=\frac{\bar{\chi}_{0}}{\bar{\xi}^{-2}+q^{2}-
\frac{i\omega}{\bar{\omega}_{F}}}
\end{equation}
where $\bar{\omega}_{F}=\frac{\omega_{F}}{2g_{0}}$ is a renormalization
group invariant quantity. In the ($z=2$) QC regime$^{[4]}$,
$\bar{\xi}^{2}\sim\frac{1}{T}$. Using these spin susceptibilities in (22)
and (23), we can betterly explain the current experimental data$^{[5-7]}$
of the nuclear magnetic resonance spin-lattice relaxation rate and the
spin echo decay rate about the copper spin. However, in the t-J model the
spin degree of the oxygen is completely suppressed, we cannot give a
reasonable explanation to the magnetic behavior of the oxygen spin only
from the t-J model.

The transport behavior of the normal state is determined by the
Lagrangian (15) or (17). The gauge field $\vec{a}$ (or $\vec{a}^{'}$) has
a drastically influence on the charge degree part because of the single
occupation condition of the electrons meaning a strong correlation
between the charge and spin degrees. In the mean field theory
approximation, if we use the hard-core boson to describe the charge
degree part of the system, the hard-core boson will move in a zero
background magnetic field deriving from the spin degree part if the phase
factor $\bar{\gamma}_{ij}$ gives a zero contribution
\begin{equation}
\vec{\partial}\cdot\vec{a}^{'}=2\pi<\tilde{f}^{+}_{\uparrow}
\tilde{f}_{\uparrow}-\tilde{f}^{+}_{\downarrow}\tilde{f}_{\downarrow}>=0
\end{equation}
where we take $<\tilde{f}^{+}_{\uparrow}\tilde{f}_{\uparrow}>
=<\tilde{f}^{+}_{\downarrow}\tilde{f}_{\downarrow}>$. If the phase factor
$\bar{\gamma}_{ij}$ gives a non-zero contribution to this background
magnetic field, the system has not a parity-symmetry in the mean field
theory approximation. If we use the slave fermion to describe the charge
degree part, in the mean field theory approximation the slave fermion
also moves in a zero background magnetic field deriving from the spin
degree part if the phase factor $\gamma_{ij}$ gives a zero contribution
\begin{equation}
\vec{\partial}\cdot\vec{a}=2\pi<b^{+}_{\uparrow}b_{\uparrow}-
b^{+}_{\downarrow}b_{\downarrow}>=0
\end{equation}
where we take $<b^{+}_{\uparrow}b_{\uparrow}>=<b^{+}_{\downarrow}
b_{\downarrow}>$. Therefore, the influence of the gauge field $\vec{a}$
(or $\vec{a}^{'}$) on the charge degree comes from the high order quantum
fluctuation of the spin degree part. However, there maybe exists a common
background magnetic field deriving from the spin degree part whether we
use the hard-core boson or the slave fermion to describe the charge
degree part because of the appearance of the phase factor $\gamma_{ij}$
(or $\bar{\gamma}_{ij}$). On the other hand, because of the single
occupation
condition, there exists a strong correlation betwwen the spin and charge
parts,
the mean field theory approximation is not accurate. We think that
the Lagrangians (15) and (17) are equivalent, we can use one of them to
describe the transport behavior of the normal state of the cuprate
superconducting materials. If we use the Lagrangian (17) to describe the
transport behavior of the system, the resistivity produced by the
quasiparticle-gauge fluctuation scattering$^{[9]}$ is $\rho\sim T$ in the
high temperature region. While in the low temperature region, the
hard-core nature of the slave boson is dominant, which derives from the
single occupation condition, so we should first change the slave boson
into a fermion which naturally obeys the single occupation constraint,
then we can calculate the transport property of the system, or we use the
Lagrangian (15) to calculate the transport behavior of the system.
If the phase factor $\gamma_{ij}$ gives a zero background magnetic field,
we can obtain the temperature dependence of the resistivity induced by
the quasiparticle-gauge fluctuation scattering$^{[9]}$
$\rho\sim T^{4/3}$. In the underdoping case, the current experimental
data shows that in the lower teperature region the resistivity is
$\rho\sim T^{\alpha}$, $1<\alpha<2$, in the higher teperature region the
resistivity is $\rho\sim T$. In the optimal doping case, the resistivity
is $\rho\sim T$ for $T>T_{c}$. Therefore, using the Lagrangians (15) and
(17), we can qualitatively explain the temperature dependence of the
resistivity. Because of the appearance of the gauge field $\vec{a}$ (or
$\vec{a}^{'}$), the high order quantum fluctuation of the spin degree
part gives an effective magnetic field $B=\vec{\partial}\cdot\vec{a}$
(ro $B^{'}=\vec{\partial}\cdot\vec{a}^{'}$), which will destroy the
parity-symmetry of the system, and may provide an odd-parity gauge
interaction $D_{j0}(q,\omega)=<A_{j}(q,\omega)A_{0}(-q,-\omega)>=\sigma
\varepsilon_{jk}q_{k}F(q^{2},\omega)$, introduced by the authors in
Ref.13. Using this interaction, we can qualitatively explain the unusual
temperature dependence of the Hall coefficient$^{[4,13]}$.

In conclusion, using the spin-hole coherent state representation, we have
studied the normal state property of the t-J model in the usual slave
boson and slave fermion treatment of the single occupation constraint,
and shown that we can qualitatively explain the unusually magnetic and
transport behaviors of the normal state of the cuprate superconducting
materials by the t-J model. We think that the short range
antiferromagnetic correlation induces the unusual behavior of the normal
state of the cuprate materials, so it is a reasonable approximation that we
take a long range antiferromagnetic
N\'{e}el order as a background of the spin degree part of the system, the
coupling between the charge degree and spin degree will destroy this long
range order, but the system still has the short range antiferromagnetic
order. In the undoping case, the system can be described by a non-linear
$\sigma$-model (the t-J model reduces to the Heisenberg model). In the
doping case, the coupling between the charge degree and spin degree
provides a decay term to the non-linear $\sigma$-model, which describes
the damping of the quasiparticle-hole pairing excitation to the spin wave
spectrum. Using this effective Lagrangian (11), we can betterly explain
the unusually magnetic behavior of the normal state of the cuprate
superconducting materials. However, there exist two Lagrangians (15) and
(17) to the charge degree part of the system, we think that they are
equivalent to each other. Using them we can qualitatively explain the
transport behavior of the normal state of the cuprate superconducting
materials, the temperature dependence of the resistivity is mainly
determined by the quasiparticle-gauge fluctuation scattering.

The author is particularly thankful to Prof. L. Yu and Prof. Z. B. Su for
their encouragement.

\newpage
{\bf References}
\bigskip
\begin{description}
\item [1]   S.Chakravarty, B.I.Halperin, and D.R.Nelson, Phys. Rev.
{\bf B}39, 2344(1989); S.Chakravarty, and R.Orbach, Phys. Rev. Lett. {\bf
64}, 224(1990).
\item [2]   S.Sachdev, and J.Ye, Phys. Rev. Lett. {\bf 69}, 2411(1992);
A.V.Chubukov, and S.Sachdev, Phys. Rev. Lett. {\bf 71}, 169(1993);
A.V.Chubukov, S.Sachdev, and J.Ye, Phys. Rev. {\bf B}49, 11919(1994).
\item [3]   A.Millis, H.Monien, and D.Pines, Phys. Rev. {\bf B}42, 167(1990);
T.Moriya, Y.Takahashi, and K.Ueda, J. Phys. Soc. Jpn. {\bf 59}, 2905(1990);
H.Monien, P.Monthoux, and D.Pines, Phys. Rev. {\bf B}43, 275(1991);
A.Sokol, and D.Pines, Phys. Rev. Lett. {\bf 71}, 2813(1993).
\item [4]  Y.L.Liu, and Z.B.Su, Phys. Lett. {\bf A}200, 393(1995);
Y.L.Liu, J. Phys.: Condens. Matter {\bf 7}, 3749(1995); Theory of the
normal state of the cuprate superconducting materials, Cond-Mat/9510015.
\item [5]  C.H.Pennington, and C.P.Slichter, Phys. Rev. Lett. {\bf 66},
381(1991); T.Imai, {\it et al}., Phys. Rev. {\bf B}47, 9158(1993); Phys.
Rev. Lett. {\bf 70}, 1002(1993); {\it ibid}. {\bf 71}, 1254(1993).
\item [6]  J.Rossat-Mignod, {\it et al}., Physica (Amsterdam) 185-189C,
86(1991); P.M.Gehring, {\it et al}., Phys. Rev. {\bf B}44, 2811(1991).
\item [7]  M.Takigawa, {\it et al}., Phys. Rev. {\bf B}43, 247(1991);
M.Takigawa, {\it ibid}. {\bf B}49, 4158(1994).
\item [8]   L.Ioffe, and A.Larkin, Phys. Rev. {\bf B}39, 8988(1989);
P.A.Lee, Phys. Rev. Lett. {\bf 63}, 6801(1989).
\item [9]   N.Nagaosa, and P.A.Lee, Phys. Rev. Lett. {\bf 64}, 2450(1990);
Phys. Rev. {\bf B}43, 1223(1991); {\bf B}43, 1234(1991).
\item [10]   L.Ioffe, and P.W.Wiegmann, Phys. Rev. Lett. {\bf 65}, 653(1990);
L.B.Ioffe, and G.Kotliar, Phys. Rev. {\bf B}42, 10348(1990).
\item [11]   A.Auerbach, {\it Interacting Electrons and Quantum
Magnetism}, (Spinger-Verlag, 1994).
\item [12]   F.D.M.Haldane, {\it Two-Dimensional Strongly Correlated
Electronic System}, Eds. Z.Z.Gan and Z.B.Su, (Gorden and Breach, 1988),
pp.249-261; Phys. Rev. Lett. {\bf 61}, 1029(1988).
\item [13]  L.B.Ioffe, V.Kalmeyer, and P.B.Wiegmann, Phys. Rev. {\bf B}43,
1219(1991).

\end{description}
\end{document}